\documentstyle[12pt]{article}
\begin{document}

{\LARGE \bf Scattering in highly singular} \\

{\LARGE \bf potentials} \\ \\

{\bf Elem\'{e}r E ~Rosinger} \\ \\
{\small \it Department of Mathematics \\ and Applied Mathematics} \\
{\small \it University of Pretoria} \\
{\small \it Pretoria} \\
{\small \it 0002 South Africa} \\
{\small \it eerosinger@hotmail.com} \\ \\

{\bf Abstract} \\

Recently, in Quantum Field theory, there has been an interest in scattering in highly singular
potentials. Here, solutions to the stationary Schr\"{o}dinger equation are presented when the
potential is a multiple of an arbitrary positive power of the Dirac delta distribution. The
one dimensional, and spherically symmetric three dimensional cases are dealt with. \\ \\

{\bf 1. One space dimension} \\

In Rosinger [4-6] solutions to the following stationary Schr\"{o}dinger equations are given \\

(1) $~~~ \frac{d^2}{dx^2}~ \psi ( x ) + ( k - U ( x ) )~ \psi ( x ) ~=~ 0,~~~ x \in {\bf R} $ \\

for any $k \in {\bf R}$, while the potentials $U$ are scalar multiples of arbitrary positive
powers of the Dirac delta distribution $\delta$, that is \\

(2) $~~~ U ( x ) ~=~ c~ \delta^m ( x ),~~~ x \in {\bf R} $ \\

with $c \in {\bf R}$ and $m > 0$ chosen at will. \\

The solutions obtained are of the form \\

(3) $~~~ \psi ( x ) ~=~ \begin{array}{|l}
                             ~\psi_{-}~ ( x ) ~~~\mbox{if}~~ x < 0 \\ \\
                             ~\psi_{+}~ ( x ) ~~~\mbox{if}~~ x > 0
                    \end{array} $ \\ \\

Here $\psi_{-}$ and $\psi_{+}$ are classical solutions of the corresponding potential free
Shcr\"{o}dinger equation \\

(4) $~~~ \frac{d^2}{dx^2}~ \psi ( x ) + k~ \psi ( x ) ~=~ 0,~~~ x \in {\bf R} $ \\

however, they must satisfy at $ x = 0$ the following {\it junction} conditions \\

(5) $~~~ \left ( \begin{array}{l}
                      ~\psi_{+}~ ( 0 ) \\ \\
                      \frac{d}{dx}~ \psi_{+}~ ( 0 )
             \end{array} \right ) ~=~ J ( m , c ) \left ( \begin{array}{l}
                       ~\psi_{-}~ ( 0 ) \\ \\
                       \frac{d}{dx}~ \psi_{-}~ ( 0 )
             \end{array} \right ) $ \\ \\

The 2$\times$2 junction matrix $J ( m, c )$ which determines the junction conditions (5)
depends on $m$ and $c$ in (2), but does not depend on $k$ in (1), and it is given respectively
by the following four cases \\

I) If $0 < m < 1$ and $c \in {\bf R}$, then \\

(6) $~~~ J ( m, c ) ~=~ \left ( \begin{array}{l} ~1~~~~0~ \\ \\
                                             ~0~~~~1~
                            \end{array} \right ) $ \\

II) If $m = 1$ and $c \in {\bf R}$, then \\

(7) $~~~ J ( m, c ) ~=~ \left ( \begin{array}{l} ~1~~~~0~ \\ \\
                                             ~c~~~~1~
                            \end{array} \right ) $ \\

III) If $m = 2$ and $c = - ( n~ \pi )^2$, with $n = 0, 1, 2, \ldots$, then \\

(8) $~~~ J ( m, c ) ~=~ \left ( \begin{array}{l} ~( - 1 )^n~~~~0~ \\ \\
                                             ~~0~~~~( - 1 )^n~
                            \end{array} \right ) $ \\

IV) If $2 < m < \infty$ and $- \infty < c < 0$, then \\

(9) $~~~ J ( m, c ) ~=~ \left ( \begin{array}{l} ~a~~~~0~ \\ \\
                                             ~b~~~~1~
                            \end{array} \right ) $ \\

where $a = \pm 1$, while $b \in {\bf R}$ can be taken arbitrary. \\

For other values of $m$ and $c$, and in particular, for $1 < m < 2$, the method used to
compute the junction matrix $J ( m, c )$, a method specified below, cannot lead to well
determined values. \\

The interpretation of the above solution (3) of the scattering problem (1) with potentials
given by multiples of positive powers of the Dirac delta distribution is as follows \\

I) For $0 < m < 1$ the potentials $U$ in (2) have {\it no} effect on the scattering. \\

II) For $m = 1$ the well known scattering results. \\

III) For $m = 2$ there are two cases \\

$~~~$ IIIa)~ For the discrete levels of the potentials \\

(10) $~~~ U ( x ) ~=~ - ( n~ \pi )^2~ \delta^2 ( x ),~~~ x \in {\bf R},~~ n ~=~ 1, 3, 5,
                                                                                   \ldots $ \\

\hspace{1.5cm} the wave function solutions $\psi$ in (3) simply suffer a change

\hspace{1.5cm} of sign, when scattering at $x = 0$ through the potentials (10). \\

$~~~$ IIIb)~ On the other hand, the discrete levels of the potentials \\

(11) $~~~ U ( x ) ~=~ - ( n~ \pi )^2~ \delta^2 ( x ),~~~ x \in {\bf R},~~ n ~=~ 0, 2, 4,
                                                                                   \ldots $ \\

\hspace{1.5cm} have {\it no} scattering effect on the wave function solutions $\psi$

\hspace{1.5cm} in (3). \\

IV) For $2 < m < \infty$, the scattering through the potentials \\

(12) $~~~ U ( x ) ~=~ c~ \delta^m ( x ),~~~ x \in {\bf R},~~ - \infty < c < 0 $ \\

exhibits a {\it double inderterminacy} involving two arbitrary real valued constants $a$ and
$b$, with $a$ only having two values, namely, $\pm 1$. \\

The method of computation of the junction matrix $J ( m, c )$ in the junction conditions (5)
follows from the nonlinear algebraic theory of generalized functions introduced and developed
in Rosinger [1-13], see also Mallios \& Rosinger [1,2] for latest developments. This nonlinear
theory constructs the largest possible class of differential algebras of generalized functions
which contain the Schwartz distributions, see www.ams.org/index/msc/46Fxx.html. This class of
algebras contains as a particular case the subsequently introduced Colombeau algebras, see
MR 92d:46098, or Grosser, et. al. [p. 7]. And as it turns out, calculations similar to those
used for obtaining the above expressions for the junction matrix $J ( m, c )$ in the junction
conditions (5) cannot be reproduced within the particular framework of the Colombeau
algebras. \\ \\

{\bf 2. Spherically symmetric three space dimensions} \\

As is well known, the stationary quantum scattering in a three dimensional and spherically
symmetric case with no angular momentum has the radial wave function $R$ given by \\

(13) $~~~ \frac{d}{dr}~( r^2~ \frac{d}{dr}~ R ( r ) ) + r^2~ ( k - U ( r ) )~ R ( r ) ~=~ 0,~~~
                                             0 < r < \infty $ \\

where $k \in {\bf R}$. If the potential $U$ is concentrated on the sphere of radius $a > 0$,
and is given by a multiple of a positive power of the Dirac delta distributions, that is \\

(14) $~~~ U ( r ) ~=~ c~ \delta^m ( r - a ),~~~ 0 < r < \infty $ \\

where $c \in {\bf R}$ and $m > 0$ are arbitrary, then, Flugge, the scattering problem (13),
(14) can be reduced to the one dimensional case (1), (2), with the consequent solution and
interpretation for the radial wave funtion $R$. \\

\end{document}